\documentclass[aps,prl,twocolumn,amsmath,amssymb,nofootinbib,preprintnumbers]{revtex4}

\usepackage{amsmath,latexsym,amssymb,amsfonts}
\usepackage[dvips]{graphicx,color}
\usepackage{bm}
\usepackage{gensymb}
\usepackage{ulem}
\usepackage[dvipsnames]{xcolor}

\begin{document}

\preprint{YITP-21-110}

\title{\textbf{Solving information loss paradox via Euclidean path integral}}

\author{
Pisin Chen$^{a,b,c,d}$\footnote{{\tt pisinchen{@}phys.ntu.edu.tw}},
Misao Sasaki$^{a,e,f}$\footnote{{\tt misao.sasaki{@}ipmu.jp}},
Dong-han Yeom$^{g,h}$\footnote{{\tt innocent.yeom{@}gmail.com}}
and
Junggi Yoon$^{i}$\footnote{{\tt junggiyoon{@}gmail.com}}
}

\affiliation{
$^{a}$Leung Center for Cosmology and Particle Astrophysics, National Taiwan University, Taipei 10617, Taiwan\\
$^{b}$Department of Physics, National Taiwan University, Taipei 10617, Taiwan\\
$^{c}$Graduate Institute of Astrophysics, National Taiwan University, Taipei 10617, Taiwan\\
$^{d}$Kavli Institute for Particle Astrophysics and Cosmology, SLAC National Accelerator Laboratory, Stanford University, Stanford, California 94305, USA\\
$^{e}$Kavli Institute for the Physics and Mathematics of the Universe (WPI), University of Tokyo, Chiba 277-8583, Japan\\
$^{f}$Center for Gravitational Physics, Yukawa Institute for Theoretical Physics, Kyoto University, Kyoto 606-8502, Japan\\
$^{g}$Department of Physics Education, Pusan National University, Busan 46241, Republic of Korea\\
$^{h}$Research Center for Dielectric and Advanced Matter Physics, Pusan National University, Busan 46241, Republic of Korea\\
$^{i}$Asia Pacific Center for Theoretical Physics, Pohang 37673, Republic of Korea
}

\begin{abstract}
The information loss paradox associated with black hole Hawking evaporation is an unresolved problem in modern theoretical physics. In this paper, we revisit the entanglement entropy via the Euclidean path integral (EPI) of the quantum state and allow for the branching of semi-classical histories along the Lorentzian evolution. We posit that there exist at least two histories that contribute to EPI, where one is an information-losing history while the other is information-preserving. At early times, the former dominates EPI, while at late times the latter becomes dominant. By so doing we recover the essence of the Page curve and thus the unitarity, albeit with the turning point, i.e., the Page time, much shifted toward the late time. One implication of this modified Page curve is that the entropy bound may thus be violated. We comment on the similarity and difference between our approach and that of the replica wormholes and the island conjecture.
\end{abstract}

\maketitle

\paragraph{Introduction}

The information loss paradox of black holes \cite{Hawking:1974sw} is an unresolved problem in modern theoretical physics. This paradox implies a contradiction between general relativity (GR) and local quantum field theory (QFT) \cite{Yeom:2009zp,Almheiri:2012rt}. There have been attempts to solve the paradox within GR and QFT. For example, the `soft hair' proposed by Hawking, Perry and Strominger \cite{Hawking:2016msc} invokes the BMS symmetry within GR, but it was soon argued that the soft hair cannot carry information \cite{Bousso:2017dny,Giddings:2019hjc}. The `firewall' conjecture \cite{Almheiri:2012rt}, on the other hand, attempts to solve the paradox by violating GR equivalence principle near the black hole horizon, but was argued to be problematic \cite{Chen:2015gux,Hwang:2012nn}. In order to render the black hole evaporation process unitary, it may be necessary to invoke some unknown new mechanism or a hidden sector \cite{Chen:2021fve,Hotta:2017yzk,Ong:2016iwi} that lies outside proper GR and QFT. A natural consequence is that such new element must be derived from \textit{quantum gravity}.

It is important to stress that \textit{entanglement entropy} is the physical quantity that measures the information flow from a black hole to its radiation \cite{Page:1993wv}. In this bipartite system of a black hole and its Hawking radiation, the radiation entropy will increase as the evaporation proceeds. On the other hand, the black hole entropy, known as the Bekenstein entropy, which is supposed to describe the black hole's microscopic degrees of freedom, will decrease as the black hole mass decreases. Then in the middle of the process, the two versions of the entropy coincide with each other. Page asserted that the entanglement entropy of the system is approximately the minimum of the two \cite{Page:1993df}. This curve of the evolution of the entanglement entropy, known as the Page curve, with the turning point, the Page time, occurring at a time when the black hole mass reduces to roughly half, plays an essential role in the investigation of the information loss problem.

In spite of Page’s demonstration in a quantum mechanical system, the attempt to derive the Page curve in GR has failed \cite{Hwang:2017yxp}, which is often regarded as the deficiency of the semi-classical perturbative calculations in GR. In particular, the decrease of the entanglement entropy after Page time would require non-perturbative effects in quantum gravity beyond our current understanding of GR. One possible circumvention of a full-blown quantum gravity calculation, which does not yet exist, would be via a new classical saddle point, e.g., a soliton, deduced from a valid approximation of quantum gravity, where a semi-classical tunneling around such a new saddle-point, e.g., via instantons, might be able to
capture the essence of non-perturbative quantum effects evaluated around the original saddle-point. Under this light, the Page curve would result from nothing but a transition between those two saddles. 

Under this philosophy, that there exist two stages in black hole evaporation. First, before a modified Page time, GR and QFT work well and any hidden contributions are negligible. After the modified Page time, however, the hidden contributions are no more negligible. In fact, it must be dominant at late times. Otherwise, the contributions via proper GR and QFT may erase the information. 

Inspired by the above thinking, we investigate the information loss paradox via the Euclidean path integral (EPI) approach \cite{Hartle:1983ai}. The EPI formalism is widely regarded as one of the most promising candidates of quantum gravity that can describe a non-purturbative processes \cite{Gibbons:1994cg}. 
Though not the final theory, EPI manages to capture the essence of a full-blown quantum gravity theory by dealing with the \textit{entire} wave-function, which includes not only perturbative but also non-perturbative gravity effects \cite{Chen:2018aij} via Wheeler-DeWitt equation \cite{DeWitt:1967yk}.

\paragraph{Essential conditions}

We posit that a successful solution to the black hole information loss paradox must satisfy the following two conditions during the black hole evaporation.

\noindent
1. \underline{Multi-history condition}: There exist at least two histories, say $h_{1}$ and $h_{2}$, that contribute to EPI, where $h_{1}$ is an information-losing history while $h_{2}$ is an information-preserving history \cite{Maldacena:2001kr,Hawking:2005kf,Sasaki:2014spa}. 

\noindent
2. \underline{Late-time dominance condition}: At early times, the probability $p_{1}$ of $h_{1}$ dominates EPI. At late times, the probability $p_{2}$ of $h_{2}$ becomes dominant. 

Here by "information-losing history" we mean the semi-classical history of an evaporating black hole in which the unitary evolution would be lost when the black hole has completely
evaporated. Based on these two conditions, the entanglement entropy $S_{\mathrm{ent}}$ can be approximately expressed as (we will prove this equation later)
\begin{eqnarray}
S_{\mathrm{ent}} \simeq p_{1} S_{1} + p_{2} S_{2},
\end{eqnarray}
where $S_{1,2}$ is the entanglement entropy of $h_{1,2}$, respectively. Initially, $p_{2} \ll p_{1} \simeq 1$ and eventually, $p_{1} \ll p_{2} \simeq 1$. $S_{1}$ increases monotonically or never approaches zero due to the information-losing nature of the history \cite{Hwang:2017yxp}, whereas $S_{2}$ eventually decreases to zero. Therefore, if these two conditions are satisfied, then the unitary Page curve would be explained, because the entanglement entropy eventually becomes zero again at the end. We will demonstrate that EPI can indeed deliver such a conclusion.

\paragraph{Euclidean path integral}

Let us focus on the following  EPI \cite{Hartle:1983ai}:
\begin{eqnarray}
\langle \Psi^{\text{\tiny cl}}_j | \textrm{in} \rangle = \int \mathcal{D} g \mathcal{D}\phi \; e^{-S_{\mathrm{E}}[g,\phi]},
\end{eqnarray}
where $| \textrm{in} \rangle$ is the initial state, $g$ is the metric, $\phi$ is a matter field, $S_{\mathrm{E}}$ is the Euclidean action, and we sum over all histories that connect $| \textrm{in} \rangle$ with a classical boundary $|\Psi^{\text{\tiny cl}}_j\rangle$. The final out-state can therefore be expressed as a superposition of many classical boundaries \cite{Hartle:2015bna,Chen:2015lbp}:
\begin{equation}
    | \textrm{out} \rangle= \sum_{j} a_j|\Psi^{\text{\tiny cl}}_j\rangle,
\end{equation}
where $a_{j} = \langle \Psi^{\text{\tiny cl}}_j | \textrm{in} \rangle$ is the coefficient of a specific classical future boundary.

\begin{figure}
\begin{center}
\includegraphics[scale=0.3]{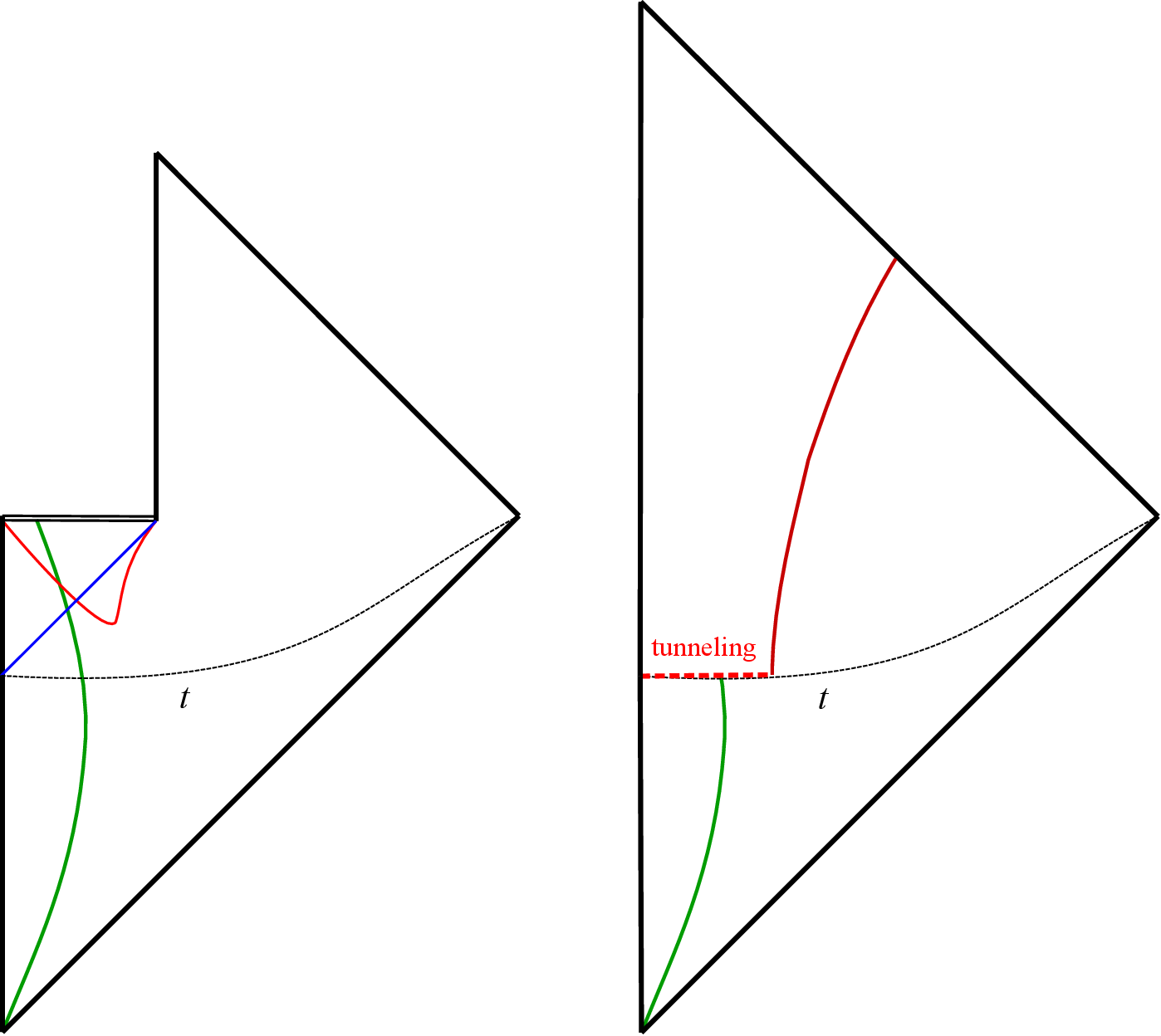}
\caption{\label{fig:conceptual2-3}Left: the causal structure of the usual semi-classical black hole, where the green curve is the trajectory of the collapsing matter, the red curve is the apparent horizon, and the blue line is the event horizon. Right: the causal structure after a quantum tunneling at the time slice $t$. After the tunneling, matter or information (red curve) is emitted and the black hole structure disappears.}
\end{center}
\end{figure}

\begin{figure}
\begin{center}
\includegraphics[scale=0.6]{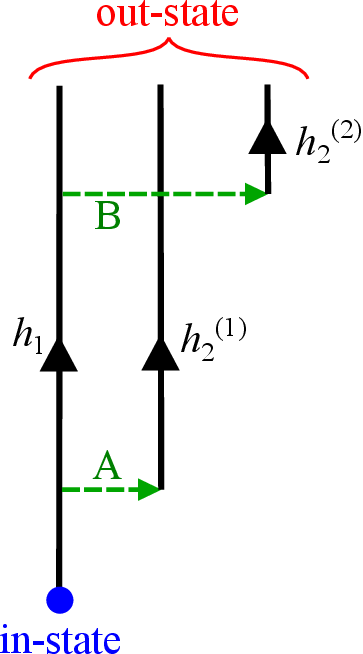}
\caption{\label{fig:conceptual4-1}Conceptual figure for interpretations. $h_{1}$ is the information-losing history, while $h_{2}^{(1,2)}$ are the information-preserving history. For any time, histories can be branched; the tunneling probability must be dominated at the late time.}
\end{center}
\end{figure}

Let $h_{1}$ be the usual history with a semi-classical black hole geometry (left of Fig.~\ref{fig:conceptual2-3}), while $h_{2}$ is an information-preserving geometry due to the non-perturbative tunneling (right of Fig.~\ref{fig:conceptual2-3}). Under normal situations the information-preserving geometry is exponentially suppressed. We will show that such a history becomes dominant at late times. Since tunneling can happen at any time, the histories will continue to branch out. Eventually, infinitely many histories appear in the out-state, while the in-state is fixed (Fig.~\ref{fig:conceptual4-1}) \cite{Hartle:2015bna,Chen:2015lbp}.

\paragraph{Thermal thin-shell model}

In order to realize the information-preserving history in the EPI approach, we introduce the thin-shell model \cite{Israel:1966rt} without black hole interiors. We consider a spacetime with spherically symmetric metric: $ds_{\pm}^{2}= - f_{\pm}(R) dT^{2} + f_{\pm}^{-1}(R) dR^{2} + R^{2} d\Omega^{2}$. Let a thin-shell be located at $r$, which follows the metric $ds^{2} = - dt^{2} + r^{2}(t) d\Omega^{2}$. The region outside the shell corresponds to $r < R$ (denoted by $+$), and that inside the shell is $R < r$ (denoted by $-$). We impose the metric ansatz for outside and inside the shell, $f_{\pm}(R) = 1 - 2M_{\pm}/R$, where $M_{+}$ and $M_{-}$ are the mass parameters of each region. We assume $M_{+} > M_{-} = 0$, and hence the inside has no black hole.

The equation of motion of the thin-shell is determined by the junction equation \cite{Israel:1966rt}:
\begin{eqnarray}\label{eq:junc}
\epsilon_{-} \sqrt{\dot{r}^{2}+f_{-}(r)} - \epsilon_{+} \sqrt{\dot{r}^{2}+f_{+}(r)} = 4\pi r \sigma(r).
\end{eqnarray}
Here, we impose $\epsilon_{\pm} = + 1$. $\sigma(r)$ is the tension parameter, where it satisfies the energy conservation relation
\begin{eqnarray}
\sigma' = -\frac{2\sigma\left(1 + w\right)}{r},
\end{eqnarray}
where $w$ is the equation of state of the shell. $w \geq -1$ is necessary for the null energy condition. In general, we assume the following form of the tension \cite{Chen:2015lbp}
\begin{eqnarray}
\sigma(r) = \sum_{i=1}^{n} \frac{\sigma_{i}}{r^{2(1+w_{i})}},
\end{eqnarray}
where $\sigma_{i}$ and $w_{i}$ are constants.

The conditions for the existence of thermal thin-shell solutions are as follows: there exists a value $r_{0}$ such that $\dot{r}|_{r=r_{0}} = 0$. This is possible, for example, if one chooses $w_{1} = 0.5$, $w_{2} = 1$, and tune $\sigma_{1,2}$. This ensures the asymptotic flatness of the spacetime, and can thus mimic causal structures of both perturbative and non-perturbative processes due to Hawking radiation.

\paragraph*{Late-time dominance condition}

For each on-shell history of EPI, the nucleation rate is $\Gamma \propto e^{-2B}$, where
\begin{eqnarray}
B = S_{\mathrm{E}}\left(\mathrm{solution}\right) - S_{\mathrm{E}}\left(\mathrm{background}\right). 
\end{eqnarray}
Due to the thermal condition, the shell is stationary in Euclidean signatures. Hence the Euclidean action is simply
\begin{eqnarray}
S_{\mathrm{E}} &=& - \int \sqrt{+g} d^{4}x \left( \frac{\mathcal{R}}{16\pi} - V \right) + \sigma \int_{\Sigma} \sqrt{+h} d^{3}x \nonumber \\
&& - \int_{\partial \mathcal{M}} \frac{\mathcal{K} - \mathcal{K}_{o}}{8\pi} \sqrt{+h} d^{3}x,\label{eq:ac}
\end{eqnarray}
where $\mathcal{R}$ is the Ricci scalar, $V$ is the vacuum energy of the matter fields, $\Sigma$ is the hypersurface of the thin-shell, and $\mathcal{K}$ and $\mathcal{K}_{o}$ are the Gibbons-Hawking boundary terms at infinity for the solution and the Minkowski background, respectively \cite{Garriga:2004nm}. 
After straightforward computations, we obtain $2B = 4 \pi M_{+}^{2}$ \cite{Chen:2017suz}.

If the shell is static, then the solution has two parts, one from the bulk integration, $4\pi M_{+}^{2}$, and the other from the boundary term at infinity, $4\pi M_{+}^{2}$. The last term in Eq.~(\ref{eq:ac}) must be cancelled between the solution of the boundary term at infinity and that of the background. Eventually, one obtains the result $2B = 4 \pi M_{+}^{2}$. 

There is an important point in this computation that we like to emphasize. In general, the Euclidean time of the background cannot be chosen freely. The Euclidean periodicity of the solution, on the other hand, is free to adjust. In principle, as long as the final boundary remains the same, one can introduce an arbitrary periodicity for the Euclidean time. One then obtains the following factorization:
\begin{eqnarray}
2B &=& n \left( (\mathrm{bulk\; term\; of\; solution}) \right. \nonumber \\
&& + \left. (\mathrm{boundary\; term\; of\; solution}) \right) \nonumber \\
&& - (\mathrm{boundary\; term\; of\; background}),
\end{eqnarray}
where $n \geq 1$. Interestingly, $n = 1$ is the most dominant contribution (one can thus ignore higher terms in general), but the subdominant contributions become important as $M_{+}$ decreases. The tunneling probability is therefore
\begin{eqnarray}
\sum_{n=1}^{\infty} e^{-S(2n-1)} = \frac{1}{e^{S} - e^{-S}},
\end{eqnarray}
where $S = 4\pi M_{+}^{2}$.

\begin{figure}
\begin{center}
\includegraphics[scale=0.8]{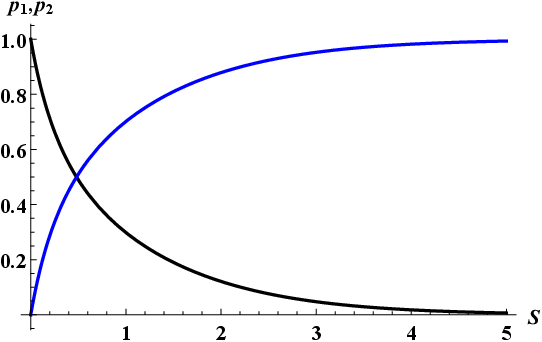}
\caption{\label{fig:prob}Probabilities of the semi-classical history ($p_{1}$, blue curve) and histories with thermal-shell emission ($p_{2}$, black curve) as a function of the entropy $S$. For large $S$, $p_{1}$ is dominated; however, there exists a golden cross between two probabilities, and eventually, $p_{2}$ is dominated.}
\end{center}
\end{figure}

If one considers only two histories, where one is the semi-classical black hole and the other is the black hole with a thermal-shell emission, then the probability of the history, $p_{1}$ and $p_{2}$, respectively, are as follows:
\begin{eqnarray}
p_{1} = \frac{e^{S} - e^{-S}}{1 + e^{S} - e^{-S}}, \;\;\;\; p_{2} = \frac{1}{1 + e^{S} - e^{-S}}.
\end{eqnarray}
Interestingly, there is a golden-cross point near $S \simeq 1$ 
(Fig.~\ref{fig:prob}). 
Such a golden-cross is actually generic. That is, even if one does not invoke multiple Euclidean time-periods, one can still obtain the same qualitative result of such a crossing, e.g., by assuming the time-accumulation, etc.

\paragraph*{Modification of the Page curve}

A generic quantum state with two classical histories can be described as follows: $| \psi \rangle = c_{1} | \psi_{1} \rangle + c_{2} | \psi_{2} \rangle$, where $1$ and $2$ denote two different histories and $c_{1,2}$ are complex coefficients. The total density matrix $\rho$ is
\begin{eqnarray}
\rho = \begin{pmatrix} |c_{1}|^{2} & c_{1}^{*} c_{2} \\ c_{1} c_{2}^{*} & |c_{2}|^{2} \end{pmatrix}.
\end{eqnarray}
We assume that after averaging over a long time, the off-diagonal terms will become less dominated; this is in accordance with the decoherence condition. With this assumption, one can write
\begin{eqnarray}
\rho \simeq \begin{pmatrix} p_{1} & 0 \\ 0 & p_{2} \end{pmatrix},
\end{eqnarray}
where $p_{1} = |c_{1}|^{2}$ and $p_{2} = |c_{2}|^{2}$.

\begin{figure}
\begin{center}
\includegraphics[scale=0.8]{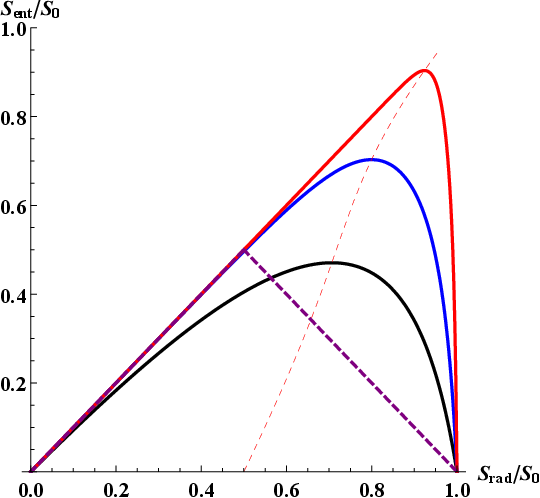}
\caption{\label{fig:ent}Entropy of emitted radiation $S_{\mathrm{rad}}/S_{0}\equiv 1 - S/S_{0}$ vs. entanglement entropy $S_{\mathrm{ent}}/S_{0}$, where $S_{0} = 3$ (black), $10$ (blue), $50$ (red) are initial black hole entropies, respectively. The purple dashed curve is the usually expected Page curve, while there exists a region where the curve is outside the purple dashed triangle region. The thin red dashed curve is the location of the Page time, i.e., $dS_{\mathrm{ent}}/dS_{\mathrm{rad}} = 0$.}
\end{center}
\end{figure}

Now let us assume that for each $|\psi_{1}\rangle$ and $|\psi_{2}\rangle$, one can define subsystems $A$ and $B$, respectively. Physically, $A$ corresponds to the subsystem outside the horizon, while $B$ is that inside the horizon. (If there is no horizon, then $B$ is empty.) One can then write
\begin{eqnarray}
| \psi_{\nu} \rangle &=& \sum_{i} a_{\nu,i} | A^{(\nu)}_{i} \rangle | B^{(\nu)}_{i} \rangle\qquad(\nu=1,2)\, .
\end{eqnarray}
The entanglement entropy should follow the Shannon information formula:
\begin{eqnarray}
S\left(A | B\right) &=& \sum_{\nu,i} p_{\nu} |a_{\nu,i}|^{2} \log p_{\nu} |a_{\nu,i}|^{2} \\
&=& p_{1} S_{1} \left(A^{(1)} | B^{(1)}\right) + p_{2} S_{2} \left(A^{(2)} | B^{(2)}\right) \nonumber \\
&& + p_{1} \log p_{1} + p_{2} \log p_{2},
\end{eqnarray}
where the last two terms are negligible if the number of degrees of freedom of the system is much greater than $2$. This proves $S_{\mathrm{ent}} \simeq p_{1} S_{1} + p_{2} S_{2}$.

We can finally coin the entanglement entropy as a function of $S$ as
\begin{eqnarray}
S_{\mathrm{ent}} &=& \left( S_{0} - S \right) \times p_{1} + 0 \times p_{2} \\
&=& \left( S_{0} - S \right) \left(\frac{e^{S} - e^{-S}}{1 + e^{S} - e^{-S}}\right),
\end{eqnarray}
where $S_{0}$ is the initial entropy of the black hole. Here, for simplicity we assume that the entanglement entropy $p_{1}$ monotonically increases, while the entanglement entropy $p_{2}$ goes to zero (because there is no black hole at the end, Fig.~\ref{fig:ent}). This preserves the unitary of the black hole evolution. 

Two salient feature of our new result is at hand. First, the Page time (thin red dashed curve in Fig.~\ref{fig:ent}) deviates from Page's original value. In our theory it is located much later than the half-way point of the radiation entropy. Second, at late times, the entanglement entropy can become much greater than what expected in the conventional Page curve (purple dashed curve in Fig.~\ref{fig:ent}). This implies that there exists a situation where the entanglement entropy is greater than that of the areal entropy. Thus the proposed monster state or the remnant-like state \cite{Chen:2014jwq,Chen:2021fve} seems to be realized. However, this is not so surprising if the total number of states varies as time goes on.

\paragraph{Conclusion}

In this paper, we demonstrate the essence of the Page curve through the EPI approach. This approach can explain the existence of the information-preserving history, as well as the late time dominance condition. The original Page curve, however, is modified. One implication is that the entropy bound may be violated and at the end of the evaporation a monster or remnant state \cite{Chen:2014jwq} would seem permissible. 

At this point, it is worthwhile to compare our result with the recent remarkable developments in string theory about the information loss paradox. The basic idea stems from the gravitational fine-grained entropy that can be computed by quantum extremal surfaces~(QES)~\cite{Engelhardt:2014gca}, i.e., the global minimum among saddles upon extremizing the generalized entropy.
It was shown \cite{Penington:2019npb,Almheiri:2019psf,Almheiri:2019hni,Bak:2020enw} that in addition to the conventional saddle, which reproduces Hawking's result, there exists another saddle that induces an \textit{island} region inside the black hole horizon. This second saddle with an island dominates the black hole evolution after the Page time and leads to the Page curve. It has been shown that such a contribution from the new saddle can be reproduced by the replica wormhole~\cite{Penington:2019kki,Almheiri:2019qdq}.

Although our interpretation of the thin-shell tunneling shares common features with the island conjecture and the replica wormhole, there exists a clear difference. The replica wormhole is based on Euclidean path integral whereas our approach deals with branching of semi-classical histories along the Lorentzian evolution. In this sense, our interpretation of thin-shell tunneling may be more closely connected with the recent studies in the context of real-time gravitational replica wormholes \cite{Colin-Ellerin:2020mva,Colin-Ellerin:2021jev} for the generalization of the island conjecture and the replica wormholes with baby universes~\cite{Marolf:2020xie,Giddings:2020yes,Marolf:2020rpm}. More detailed comparisons of our approach with the replica wormhole approach, either Euclidean or Lorentzian, are left for future investigations.


\section*{Acknowledgment} PC is supported by Taiwan's Ministry of Science and Technology (MOST) and the Leung Center for Cosmology and Particle Astrophysics (LeCosPA), National Taiwan University. MS is supported in part by JSPS KAKENHI grant Nos. 19H01895, 20H04727, and 20H05853. DY is supported by the National Research Foundation of Korea (Grant no.: 2021R1C1C1008622, 2021R1A4A5031460). JY is supported by the National Research Foundation of Korea (Grant no.: 2019R1F1A1045971). JY is also supported by an appointment to the JRG Program at APCTP through the Science and Technology Promotion Fund, the Lottery Fund of the Korean government, and by Gyeongsangbuk-do and Pohang-si.

\end{document}